\documentclass[twocolumn,english,conference]{IEEEtran}
\usepackage[english]{babel}
\usepackage[T1]{fontenc}
\usepackage[utf8]{inputenc}
\usepackage{float}
\usepackage{amsmath}
\usepackage{amssymb}
\usepackage{amsthm}
\usepackage{graphicx}
\usepackage{epstopdf}
\usepackage{subfigure}
\usepackage{babel}
\usepackage{indentfirst}
\usepackage{ifpdf}
\usepackage{cite}
\usepackage{stfloats}
\usepackage{cite}
\usepackage{bm}
\usepackage{mathtools}
\usepackage{multirow}
\usepackage[keeplastbox]{flushend} 
\makeatletter

\providecommand{\tabularnewline}{\\}
\floatstyle{ruled}
\newfloat{algorithm}{tbp}{loa}
\providecommand{\algorithmname}{Algorithm}
\floatname{algorithm}{\protect\algorithmname}

\IEEEoverridecommandlockouts
\usepackage{ifpdf}
\usepackage{cite}
\usepackage{babel}
\usepackage{flushend}
\ifCLASSOPTIONcompsoc
\else
\fi

\makeatother

\newtheorem{remark}{\hspace{1em}\it Remark}

\begin{document}
\renewcommand{\figurename}{Fig.}
\title{OTFS Based Receiver Scheme With Multi-Antennas in High-Mobility V2X Systems}

\author{
\IEEEauthorblockN{Junqiang Cheng\IEEEauthorrefmark{2}, Chenglu Jia\IEEEauthorrefmark{2}, Hui Gao\IEEEauthorrefmark{2}, Wenjun Xu\IEEEauthorrefmark{3}, and Zhisong Bie\IEEEauthorrefmark{3}}
\IEEEauthorblockA{\IEEEauthorrefmark{2}Key Lab of Trustworthy Distributed Computing and Service, Ministry of Education}
\IEEEauthorblockA{\IEEEauthorrefmark{3}Key Lab of Universal Wireless Communications, Ministry of Education\\
Beijing University of Posts and Telecommunications, Beijing, China, 100876\\
Email: \{jqcheng, chenglujia, huigao, wjxu, zhisongbie\}@bupt.edu.cn}
}
\maketitle

\begin{abstract}
Vehicle-to-everything (V2X) is considered as one of the most important applications of future wireless communication networks. However, the Doppler effect caused by the vehicle mobility may seriously deteriorate the performance of the vehicular communication links, especially when the channels exhibit a large number of Doppler frequency offsets (DFOs). Orthogonal time frequency space (OTFS) is a new waveform designed in the delay-Doppler domain, and can effectively convert a doubly dispersive channel into an almost non-fading channel, which makes it very attractive for V2X communications. In this paper, we design a novel OTFS based receiver with multi-antennas to deal with the high-mobility challenges in V2X systems. We show that the multiple DFOs associated with multipaths can be separated with the high-spatial resolution provided by multi-antennas, which leads to an enhanced sparsity of the OTFS channel in the delay-Doppler domain and bears a potential to reduce the complexity of the message passing (MP) detection algorithm. Based on this observation, we further propose a joint MP-maximum ration combining (MRC) iterative detection for OTFS, where the integration of MRC significantly improves the convergence performance of the iteration and gains an excellent system error performance. Finally, we provide numerical simulation results to corroborate the superiorities of the proposed scheme.
\end{abstract}

\begin{IEEEkeywords}
High-mobility, OTFS, multi-antennas, beamforming, message passing.
\end{IEEEkeywords}

\section{Introduction}
As a key component of future wireless communication networks, vehicle-to-everything (V2X) communications aim to satisfy high demands in terms of the number of connected devices, data rates, latency and reliability \cite{shah20185g}. These requirements, especially when the high-mobility of vehicles is considered, make the V2X communications very challenging. Orthogonal frequency division multiplexing (OFDM) modulation is believed to be the most promising communication technology for vehicle communications \cite{cheng2015d2d}. However, the time-variant channel caused by the high-mobility in V2X communications results in significant inter-carrier interference (ICI), which seriously degrades the performance of OFDM systems \cite{jakes1994microwave}. Orthogonal time frequency space (OTFS) modulation, which was originally proposed in \cite{hadani2017otfs}, has been demonstrated to achieve a significant performance improvement over OFDM in doubly dispersive channels \cite{ramachandran2018mimo, ding2019robust, shen2019channel, ding2019otfs}. A key hallmark of OTFS is that it can effectively convert a doubly dispersive channel into an almost non-fading channel in the delay-Doppler domain \cite{hadani2018arXiv}, and consequently all symbols in a frame experience a relatively stable channel, which makes it very attractive for V2X communications with high-mobility.

The equalization scheme is critical to the performance of practical OTFS systems. Compared with the linear equalizations \cite{cheng2019low}, \cite{xu2019low}, non-linear equalizations are more suitable for OTFS for the reason that they are able to exploit the full channel diversity over both time and frequency \cite{hadani2018arXiv}, \cite{long2019low}. Message passing (MP) algorithm \cite{raviteja2018interference} is the most studied non-linear equalization scheme, which is suitable for OTFS systems given the  assumption of the channel sparsity. However, the channel sparsity level may not be well guaranteed in some V2X scenarios with rich reflectors, such as urban districts, mountain areas, and tunnels. The multipaths between the transceivers in such scenarios result in a large number of Doppler frequency offsets (DFOs), which reduce the channel sparsity and increase the complexity of the MP algorithm. Fortunately, considering that DFOs are closely associated with the angle of arrivals (AoAs) or angle of departures (AoDs) of multipaths, some pioneer works have attempted to separate the DFOs from spatial domain by exploiting the large antenna array \cite{chizhik2004slowing, guo2017high, guo2019high}, which also bears a potential to enhance the OTFS channel sparsity in the delay-Doppler domain and reduce the complexity of the MP algorithm.

Motivated by the above observations, in this paper, we propose a novel OTFS based receiver scheme with multi-antennas in high-mobility V2X systems. More specifically, we consider a high-mobility vehicle to base station (BS) uplink transmission scenario with rich reflectors, and the BS is equipped with a uniform linear array (ULA). Thanks to the high-spatial resolution provided by the ULA, the multiple DFOs associated with multipaths can be approximately separated after a beamforming network designed in this paper, which guarantees the channel sparsity in the delay-Doppler domain even in the presence of rich reflectors. Then, different from all precious works based on OFDM \cite{guo2017high}, \cite{guo2019high}, we adopt OTFS based receiver to effectively combat the residual DFOs after the beamforming network and gain a much better system error performance. Moreover, empowered by the channel sparsity after the beamforming network, the MP algorithm can be conducted at a low complexity. Based on it, we further propose a joint MP-maximum ration combining (MRC) iterative detection for OTFS, where the integration of the MRC significantly improves the convergence performance of the iteration and also obtains an excellent multi-antenna diversity. Numerical simulation results are provided to show the advantages of the proposed scheme.

The rest of this paper is organized as follows. In Section \uppercase\expandafter{\romannumeral2}, the system model is introduced, including the time-variant multipath channel model and the signal model of OTFS system. In Section \uppercase\expandafter{\romannumeral3}, an OTFS based V2X receiver design with multi-antennas is proposed, including the beamforming network design and the joint MP-MRC iterative algorithm. Simulation results are presented in Section \uppercase\expandafter{\romannumeral4}, while conclusions are drawn in Section \uppercase\expandafter{\romannumeral5}.

{\emph{Notation:} Matrix, vector and scalar are denoted by $\mathbf{A}$, $\mathbf{a}$ and $a$, respectively. $\mathbf{I}_{M}$, $\mathbf{F}_{M}$ and $\mathbf{F}_{M}^{H}$ are the $M$-point identity matrix, DFT matrix and IDFT matrix, respectively. The function $\mathrm{vec}\left(\mathbf{A}\right)$ denotes the columns-wise matrix vectorization. The superscripts $\left(\cdot\right)^{*}$ and $\left(\cdot\right)^{H}$ indicate conjugate and conjugate transpose operations, respectively. Finally, the operator $\otimes$ represents Kronecker product operation.}

\section{System Model}
\begin{figure}
  \centering
  \includegraphics[scale=0.4805]{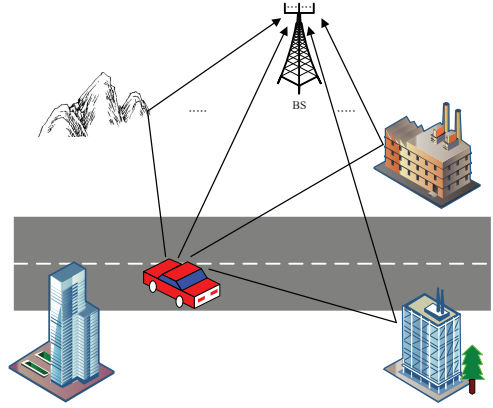}
  \caption{High-mobility vehicle to base station uplink transmission with rich reflectors.}
  \label{scenario}
\end{figure}

Consider the scenarios of high-mobility uplink vehicle communications, as shown in Fig. \ref{scenario}, where the high-speed mobile vehicle transmits signal to the BS through multipaths caused by various reflectors, and the BS is equipped with the ULA. In this section, we give the corresponding time-variant multipath channel model in such scenarios and the signal model of OTFS system in a vectorized form.

\subsection{Time-Variant Multipath Channel Model}
Assume that the ULA with $N_{r}$ antennas is configured at the BS, then the baseband time-variant multipath channel from the vehicle to the $a$-th receive antenna can be modeled as $N_{tap}$ taps with different delays
\begin{equation}
\begin{aligned}
h_{a}\left(n,n^{'}\right)=\sum_{i=1}^{N_{tap}}g_{a,i}\left(n\right)\delta\left(n^{'}-d_{i}\right),
\label{eq_channel1}
\end{aligned}
\end{equation}
where $d_{i}$ is the relative delay of the $i$-th tap. $g_{a,i}(n)$ is the corresponding complex amplitude for the $i$-th tap at the $a$-th antenna. To characterize the scenarios with rich reflectors, as in \cite{guo2019high}, each tap is comprised of $N_{q}$ propagation paths. Then the equivalent model for $g_{a,i}\left(n\right)$ can be expressed as
\begin{equation}
\begin{aligned}
g_{a,i}(n)=\sum_{q=1}^{N_{q}}\alpha_{i,q}e^{j\left[2\pi f_{i,q}nT_{s}+\phi_{a}(\theta_{i,q})\right]},
\end{aligned}
\label{eq_channel2}
\end{equation}
where $T_{s}$ is the sampling interval. $\alpha_{i,q}$, $f_{i,q}$ and $\theta_{i,q}$ denote the random complex path gain, the DFO and the AoA associated with the $q$-th path in the $i$-th tap, respectively. Further, the DFO of the $q$-th path in the $i$-th tap can be denoted as $f_{i,q}=vf_{c}cos(\beta_{i,q})/c$ with $v$, $f_{c}$ and $c$ being the speed of the vehicle, the carrier frequency and the speed of light, respectively. And $\beta_{i,q}$ is randomly distributed between $0$ and $\pi$, which is related to the moving direction of the vehicle and the AoD of the $q$-th path in the $i$-th tap. Besides, $\phi_{a}(\theta_{i,q})$ in (\ref{eq_channel2}) denotes the phase shift at the $a$-th antenna. Since ULA is adopted in this paper, the steering vector for the whole antenna array at the AoA $\theta_{i,q}$ can be expressed as
\begin{equation}
\begin{aligned}
\mathbf{a}\left(\theta_{i,q}\right)\!=\!\left[1,e^{j(2\pi/\lambda)d\mathrm{sin}(\theta_{i,q})},\cdots\!,e^{j(N_{r}-1)(2\pi/\lambda)d\mathrm{sin}(\theta_{i,q})}\right]^{T},
\end{aligned}
\label{eq_steering}
\end{equation}
where $\lambda$ is the wavelength of the carrier wave, and $d$ is the distance between antenna elements.

In a high-mobility V2X scenario, $g_{a,i}\left(n\right)$ varies with the time index $n$ due to the DFOs. Besides, when there are rich reflectors in the communication scenarios, $N_{q}$ tends to be very large and thus the channel model in (\ref{eq_channel1}) and (\ref{eq_channel2}) coincides with the classical Jakes' channel model \cite{jakes1994microwave}.

\subsection{Signal Model of OTFS System}
As presented in \cite{hadani2017otfs}, \cite{shen2019channel, ding2019otfs, hadani2018arXiv} and \cite{rezazadehreyhani2018analysis}, OTFS system can be implemented as a pre- and post-processing enhanced OFDM system. Let $\mathbf{X}\in\mathbb{C}^{M\times N}$ denotes the two-dimensional OTFS symbols in the delay-Doppler domain, where $M$ and $N$ are the numbers of resource units along the delay domain and Doppler domain. Then the OTFS symbols $\mathbf{X}$ can be transformed into the time domain through the OFDM based OTFS modulation,
\begin{equation}
\begin{aligned}
\mathbf{s}=\overset{{\scriptstyle \mathrm{Add\;CP}}}{\overbrace{\left(\mathbf{I}_{N}\otimes\mathbf{A}_{CP}\right)}}\underset{{\scriptstyle \mathrm{IDFT}}}{\underbrace{\left(\mathbf{I}_{N}\otimes\mathbf{F}_{M}^{H}\right)}}\overset{{\scriptstyle \mathrm{Window}}}{\overbrace{\mathbf{W}_{T}}}\underset{{\scriptstyle \mathrm{ISFFT}}}{\underbrace{\left(\mathbf{F}_{N}^{H}\otimes\mathbf{F}_{M}\right)}}\mathbf{x},
\end{aligned}
\label{eq_s}
\end{equation}
where $\mathbf{s}\!\in\!\mathbb{C}^{(M+N_{CP})N\times1}$ is the time domain symbols with $N_{CP}$ being the length of cyclic prefix (CP), and $\mathbf{x}\!=\mathrm{vec}\left(\mathbf{X}\right)\in\!\mathbb{C}^{MN\times 1}$ is the vectorized OTFS symbols in the delay-Doppler domain. Here, \emph{inverse symplectic finite Fourier transform} (ISFFT), together with the transmit windowing matrix, $\mathbf{W}_{T}\in\mathbb{C}^{MN\times MN}$, forms the OTFS pre-processing block. Moreover, the inverse discrete Fourier transform (IDFT) operation and CP addition operation constitute OFDM modulation, where $\mathbf{A}_{CP}\in\mathbb{C}^{(M+N_{CP})\times M}$ is the CP addition matrix.

After passing over the time-variant multipath channel described in (\ref{eq_channel1}) and (\ref{eq_channel2}), the $n$-th element of the received signal $\mathbf{r}_{a}\in\mathbb{C}^{(M+N_{CP})N\times1}$, $a=1,\cdots,N_{r}$, at the $a$-th antenna is given by
\begin{equation}
\begin{aligned}
r_{a}\left(n\right)=\sum_{i=1}^{N_{tap}}g_{a,i}\left(n\right)s\left(n-d_{i}\right)+z_{a}\left(n\right),
\end{aligned}
\label{eq_r1}
\end{equation}
where $z_{a}\left(n\right)$ is the additive noise at the $a$-th receive antenna. Next, we can obtain the received signal matrix on the whole antenna array as $\mathbf{R}\!=\!\left[\mathbf{r}_{1},\mathbf{r}_{2},\cdots\!,\mathbf{r}_{N_{r}}\right]$, which is then processed by the OTFS based receiver as detailed in the next section.

\section{OTFS Based Receiver Scheme With Multi-Antennas}

\begin{figure}
  \centering
  \includegraphics[scale=0.435]{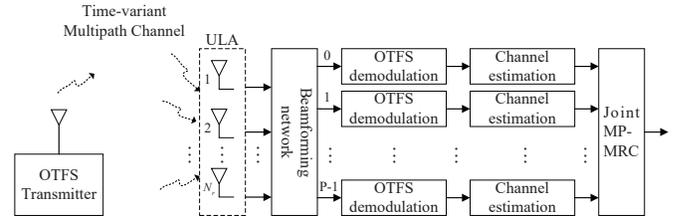}
  \caption{Diagram of the OTFS based receiver scheme with multi-antennas.}
  \label{receiver}
\end{figure}

In this section, we design an OTFS based receiver with multi-antennas. As shown in Fig. \ref{receiver}, the received signals first pass through a beamforming network, and then produce $P$ parallel beamforming branches. Next, the time domain symbols in each beamforming branch go through the OTFS demodulation to obtain the corresponding symbols in the delay-Doppler domain. Thanks to the equivalent channel of each beamforming branch enjoys an enhanced sparsity in the delay-Doppler domain, the complexity of the MP algorithm is thus reduced. Based on it, we further propose a joint MP-MRC iterative algorithm for OTFS, where the integration of the MRC significantly improves the convergence performance of the iteration and also obtains an excellent diversity gain.

\subsection{Beamforming Network}
As discussed previously, since the multiple DFOs are closely associated with the multipaths, we can approximately separate DFOs in spatial domain with the high-spatial resolution provided by the antenna array. To this end, the goal of the beamforming is to maintain the signal from only one desired direction while suppressing the signals from the other directions. Motivated by \cite{guo2019high}, the matched filter (MF) beamformer is adopted in this paper. Then, the weight vector for direction $\theta_{p}$, $p=0,\cdots,P-1$ is determined by the steering vector in (\ref{eq_steering}), that is
\begin{equation}
\mathbf{w}\left(\theta_{p}\right)=\mathbf{a}\left(\theta_{p}\right)/N_{r}.
\label{weight}
\end{equation}

Next, the received signal after beamforming network towards the direction $\theta_{p}$ can be expressed as
\begin{equation}
\mathbf{\tilde{r}}\left(\theta_{p}\right)=\mathbf{R}\cdot\mathbf{w}^{\ast}\left(\theta_{p}\right).
\label{r_w}
\end{equation}

Then, the time domain signal $\mathbf{\tilde{r}}\left(\theta_{p}\right)$ on the $p$-th beamforming branch is transformed into the delay-Doppler domain through OTFS demodulation, which consists of a traditional OFDM demodulation block and a post-processing block as mentioned previously, and can be formulated as
\begin{equation}
\begin{aligned}
\mathbf{y}\left(\theta_{p}\right)=\overset{{\scriptstyle \mathrm{SFFT}}}{\overbrace{\left(\mathbf{F}_{N}\otimes\mathbf{F}_{M}^{H}\right)}}\!\underset{\mathrm{{\scriptstyle Window}}}{\underbrace{\mathbf{W}_{R}}}\overset{{\scriptstyle \mathrm{DFT}}}{\overbrace{\left(\mathbf{I}_{N}\otimes\mathbf{F}_{M}\right)}}\underset{{\scriptstyle \mathrm{Remove\;CP}}}{\underbrace{\left(\mathbf{I}_{N}\otimes\mathbf{R}_{CP}\right)}}\tilde{\mathbf{r}}\left(\theta_{p}\right),
\end{aligned}
\label{eq_yDD}
\end{equation}
where the CP removal operation and discrete Fourier transform (DFT) operation constitute OFDM demodulation, and $\mathbf{R}_{CP}\in\mathbb{C}^{M\times(M+N_{CP})}$ is the CP removal matrix. Besides, the receive windowing matrix $\mathbf{W}_{R}\in\mathbb{C}^{MN\times MN}$, together with \emph{symplectic finite Fourier transform} (SFFT) operation forms the OTFS post-processing block.

After channel estimation, the signal $\mathbf{y}\left(\theta_{p}\right)$ on each beamforming branch is passed through a carefully designed joint MP-MRC iterative algorithm to eliminate the interferences and also obtain the multi-antenna diversity with an accelerated convergence rate as will be detailed later.

\begin{remark}
\emph{Note that due to the limited resolution of the spatial beam in the realistic communication systems, some DFOs might not be separated after the beamforming network. Therefore, the received symbols on each beamforming branch still suffer from the residual DFOs, which motivates the utilization of OTFS system instead of OFDM system to effectively combat the Doppler effect in this paper.}
\end{remark}

\subsection{Joint Message Passing-Maximum Ratio Combining Iterative Algorithm}
An important attribute of the delay-Doppler channel representation is its sparsity \cite{hadani2018arXiv}, especially when there are only a small number of paths, with varying delay and Doppler values. Consequently, the MP detection algorithm can be modeled based on a sparsely-connected factor graph between the variable nodes and the observation nodes. However, when there are a large number of paths between the transceivers as discussed in this paper, the sparsity level of the channel may not be well guaranteed, which further increases the complexity of the MP algorithm. Therefore, we propose to separate the DFOs in the spatial domain with a beamforming network designed at the receiver, which guarantees the equivalent channel on each beamforming branch enjoys an enhanced sparsity as simulated in the next section, and hence reduces the complexity of the MP algorithm. Based on it, in order to obtain the best diversity performance and accelerate the convergence rate of the MP algorithm, we further design a joint MP-MRC iterative algorithm for OTFS receiver as detailed below.

The received symbols after OTFS demodulation on $p$-th beamforming branch in (\ref{eq_yDD}) can be formulated in a vectorized form as \cite{shen2019channel}
\begin{equation}
\mathbf{y}\left(\theta_{p}\right)=\mathbf{H}\left(\theta_{p}\right)\mathbf{x}+\mathbf{z}\left(\theta_{p}\right),
\label{eq_system}
\end{equation}
where $\mathbf{x},\mathbf{y}\left(\theta_{p}\right), \mathbf{z}\left(\theta_{p}\right)\in\mathbb{C}^{MN\times1}$ and $\mathbf{H}\left(\theta_{p}\right)\in\mathbb{C}^{MN\times MN}$, $p=0,\cdots,P-1$ are the transmitted symbols, received symbols, noise term and the equivalent channel matrix represented in the delay-Doppler domain on the $p$-th branch, respectively.

Let $\Phi_{pa}$ and $\Psi_{pb}$ denote the sets of non-zero positions in the $a$-th row and $b$-th column of $\mathbf{H}\left(\theta_{p}\right)$, respectively. Based on (\ref{eq_system}), the MP algorithm can be modeled as a sparsely-connected factor graph with $MN$ variable nodes corresponding to the elements in $\mathbf{x}$ and $MN$ observation nodes corresponding to the elements in $\mathbf{y}\left(\theta_{p}\right)$. More specifically, each observation node $y\left(\theta_{p}\right)_{d}$, $d=0,\cdots,MN-1$ is connected to the set of variable nodes $\left\{ x_{e},e\in\Phi_{pa}\right\} $ and each variable node $x_{c}$, $c=0,\cdots,MN-1$ is connected to the set of observation nodes $\left\{ y\left(\theta_{p}\right)_{e},e\in\Psi_{pb}\right\} $. Then, the \emph{maximum a posteriori} (MAP) detection rule for estimating the transmitted symbols of (\ref{eq_system}) can be expressed as \cite{raviteja2018interference}
\begin{equation}
\begin{aligned}
\mathbf{\hat{x}}=\underset{\mathbf{x}\in\mathbb{A}^{MN}}{\mathrm{arg\:max}}\;\mathrm{Pr}\left(\mathbf{x}|\mathbf{y}\left(\theta_{p}\right),\mathbf{H}\left(\theta_{p}\right)\right),
\end{aligned}
\label{eq_map}
\end{equation}
where $\mathbb{A}$ is the modulation constellation, and $\mathbf{\hat{x}}$ is the estimation of the transmitted symbols on the $p$-th beamforming branch. Obviously, the complexity of the above equation is exponential in $MN$. Therefore, the symbol-by-symbol MAP detection rule for $c=0,\cdots,MN-1$ is considered\cite{raviteja2018interference}
\begin{equation}
\begin{aligned}
\hat{x}_{c}&=\underset{a_{j}\in\mathbb{A}}{\mathrm{arg\:max}}\;\mathrm{Pr}\left(x_{c}=a_{j}\mid\mathbf{y}\left(\theta_{p}\right),\mathbf{H}\left(\theta_{p}\right)\right)\\
&=\underset{a_{j}\in\mathbb{A}}{\mathrm{arg\:max}}\frac{1}{\mid\mathbb{A\mid}}\mathrm{Pr}\left(\mathbf{y}\left(\theta_{p}\right)\mid x_{c}=a_{j},\mathbf{H}\left(\theta_{p}\right)\right)\\
&\approx\underset{a_{j}\in\mathbb{A}}{\mathrm{arg\:max}}\prod_{e\in\Psi_{pb}}\mathrm{Pr}\left(y\left(\theta_{p}\right)_{e}|x_{c}=a_{j},\mathbf{H}\left(\theta_{p}\right)\right).
\end{aligned}
\label{eq_sym_map}
\end{equation}
Here, the transmitted symbols are assumed to be equally likely and the components of $\mathbf{y}\left(\theta_{p}\right)$ are approximately independent for a given $x_{c}$ due to the enhanced sparsity of $\mathbf{H}\left(\theta_{p}\right)$. The approximate symbol-by-symbol MAP detection in (\ref{eq_sym_map}) can be solved using the MP algorithm, which is used for a single-input single-output (SISO) system in \cite{raviteja2018interference}. In this paper, empowered by the enhanced channel sparsity after the beamforming network, we further propose a joint MP-MRC iterative algorithm to gain the multi-antenna diversity and also improve the convergence performance of the MP algorithm.

The essentials of the joint MP-MRC algorithm lie in three steps in each iteration. First, each observation node passes the mean and variance of the interference terms to the connected variable nodes $x_{e},e\in\Phi_{pa}$. Second, each variable node updates the probability mass function (pmf) $\mathbf{p}_{c,e}=\left\{ p_{c,e}\left(a_{j}\right)\mid a_{j}\in\mathbb{A}\right\}$ of the alphabet symbols in $\mathbb{A}$, and then passes it back to the connected observation nodes $y\left(\theta_{p}\right)_{e},e\in\Psi_{pb}$. Third, the \emph{joint convergence indicator} of all beamforming branches is calculated in the MRC fashion after each iteration. Finally, when the \emph{joint convergence indicator} converges, the soft output is computed for each transmitted symbol, and then the final decision is made.

The joint MP-MRC iterative algorithm operates as follows:
\begin{enumerate}
  \item \textbf{Inputs}: $\mathbf{y}\left(\theta_{p}\right)$, $\mathbf{H}\left(\theta_{p}\right)$, $p=0,\cdots,P-1$.
  \item \textbf{Initialization}: Iteration index $i=1$, pmf $\mathbf{p}_{ce}^{\left(0\right)}=1/|\mathbb{A}|$ for $c\in\left\{ 0,\cdots,MN-1\right\} $ and $e\in\Psi_{pb}$.
  \item \textbf{Messages from observation nodes to variable nodes}: The interference terms on $p$-th beamforming branch can be modeled as a Gaussian random variable $\zeta_{d,e}^{\left(p\right)}$ defined as
\begin{equation}
\begin{aligned}
y\left(\theta_{p}\right)_{d}=x_{e}H\left(\theta_{p}\right)_{d,e}+\!\underset{\zeta_{d,e}^{\left(p\right)}}{\underbrace{\sum_{t\in\Phi_{pa},t\neq e}\!\!x_{t}H\left(\theta_{p}\right)_{d,t}+\!z\left(\theta_{p}\right)_{d}}},
\end{aligned}
\label{eq_interference}
\end{equation}
where $e\in\Phi_{pa}$, and $H\left(\theta_{p}\right)_{d,e}$ represents the element of matrix $\mathbf{H}\left(\theta_{p}\right)$ at $d$-th row and $e$-th column. Then, the mean $\mu_{d,e}^{\left(p\right)\left(i\right)}$ and variance $\left(\sigma_{d,e}^{\left(p\right)\left(i\right)}\right)^{2}$ of the interference $\zeta_{d,e}^{\left(p\right)}$ can be computed as
\begin{equation}
\begin{aligned}
\mu_{d,e}^{\left(p\right)\left(i\right)}\!\!=\!\mathbb{E}\left(\zeta_{d,e}^{\left(p\right)}\right)\!=\!\!\!\sum_{t\in\Phi_{pa},t\neq e}\sum_{j=1}^{|\mathbb{A}|}p_{t,d}^{\left(p\right)\left(i\right)}\!\left(a_{j}\right)a_{j}H\!\left(\theta_{p}\right)_{d,t},
\end{aligned}
\label{eq_mean}
\end{equation}

\begin{equation}
\begin{aligned}
\left(\sigma_{d,e}^{\left(p\right)\left(i\right)}\right)^{2}&\!\!=\mathrm{Var}\left(\zeta_{d,e}^{\left(p\right)}\right)\\
&\!\!=\!\sum_{t\in\Phi_{pa},t\neq e}\!\left(\sum_{j=1}^{|\mathbb{A}|}p_{t,d}^{\left(p\right)\left(i\right)}\!\left(a_{j}\right)\left|a_{j}\right|^{2}\left|H\!\left(\theta_{p}\right)_{d,t}\right|^{2}\right.\\
&\phantom{=\;\;}
\left. \!\!-\left|\sum_{j=1}^{|\mathbb{A}|}p_{t,d}^{\left(p\right)\left(i\right)}\left(a_{j}\right)a_{j}H\left(\theta_{p}\right)_{d,t}\right|^{2}\right)+\sigma^{2}.
\end{aligned}
\label{eq_var}
\end{equation}

  \item \textbf{Messages from variable nodes to observation nodes}: The messages passed from $x_{c}$ to $y_{e}$, $e\in\Psi_{pb}$ is the pmf vector $\mathbf{p}_{c,e}$ with elements
\begin{equation}
\begin{aligned}
p_{c,e}^{\left(p\right)\left(i\right)}\left(a_{j}\right)\!=\!\Delta\cdot\tilde{p}_{c,e}^{\left(p\right)\left(i\right)}\left(a_{j}\right)+\left(1-\Delta\right)\cdot p_{c,e}^{\left(p\right)\left(i-1\right)}\left(a_{j}\right),
\end{aligned}
\label{eq_prob1}
\end{equation}
where $\Delta\in(0,1]$ is the damping factor \cite{raviteja2018interference} to affect the performance by controlling the convergence rate, and
\begin{equation}
\begin{aligned}
\tilde{p}_{c,e}^{\left(p\right)\left(i\right)}\left(a_{j}\right)
&\propto\prod_{t\in\Psi_{pb},t\neq e}\mathrm{Pr}\left(y\left(\theta_{p}\right)_{t}\mid x_{c}=a_{j},\mathbf{H}\left(\theta_{p}\right)\right)\\
&=\prod_{t\in\Psi_{pb},t\neq e}\frac{\xi^{\left(p\right)\left(i\right)}\left(t,c,j\right)}{\sum_{k=1}^{Q}\xi^{\left(p\right)\left(i\right)}\left(t,c,k\right)},
\end{aligned}
\label{eq_prob2}
\end{equation}
where
\begin{equation}
\begin{aligned}
&\xi^{\left(p\right)\left(i\right)}\left(t,c,k\right)\\
&=\mathrm{exp}\left(\frac{-\left|y\left(\theta_{p}\right)_{t}-\mu_{t,c}^{\left(p\right)\left(i\right)}-H\left(\theta_{p}\right)_{t,c}a_{k}\right|^{2}}{\left(\sigma_{t,c}^{\left(p\right)\left(i\right)}\right)^{2}}\right).
\end{aligned}
\label{eq_prob3}
\end{equation}
  \item \textbf{Joint convergence indicator}: The \emph{joint convergence indicator} $\Upsilon^{\left(i\right)}$ in the $i$-th iteration after the MRC of all beamforming branches can be computed as
\begin{equation}
\begin{aligned}
\Upsilon^{\left(i\right)}=\frac{1}{MN}\sum_{c=0}^{MN-1}\mathcal{F}\left(\underset{a_{j}\in\mathbb{A}}{\mathrm{max}}\:p_{\mathrm{NJP},c}^{\left(i\right)}\left(a_{j}\right)\geq1-\varepsilon\right),
\end{aligned}
\label{eq_converge1}
\end{equation}
for some small $\varepsilon>0$ and where the operator $\mathcal{F}\left(\cdot\right)$ is an indicator function which gives a value of 1 if the expression in the argument is true, and 0 otherwise. Moreover, $p_{\mathrm{NJP},c}^{\left(i\right)}\left(a_{j}\right)$ is the \emph{normalized joint probability} (NJP) after the MRC of all beamforming branches and is given by
\begin{equation}
\begin{aligned}
p_{\mathrm{NJP},c}^{\left(i\right)}\left(a_{j}\right)=\mathrm{nmlz}\left(\prod_{p=0,\cdots,P-1}p_{c}^{\left(p\right)\left(i\right)}\left(a_{j}\right)\right),
\end{aligned}
\label{eq_nmlzjpro1}
\end{equation}
where $\mathrm{nmlz}\left(\cdot\right)$ is the normalization operation, and
\begin{equation}
\begin{aligned}
p_{c}^{\left(p\right)\left(i\right)}\left(a_{j}\right)=\prod_{t\in\Psi_{pb}}\frac{\xi^{\left(p\right)\left(i\right)}\left(t,c,j\right)}{\sum_{k=1}^{Q}\xi^{\left(p\right)\left(i\right)}\left(t,c,k\right)}.
\end{aligned}
\label{eq_nmlzjpro2}
\end{equation}
  \item \textbf{Update decision}: If $\Upsilon^{\left(i\right)}>\Upsilon^{\left(i-1\right)}$, then the decision of the transmitted symbols is updated,
\begin{equation}
\begin{aligned}
\hat{x}_{c}=\underset{a_{j}\in\mathbb{A}}{\mathrm{arg\;max}}\;p_{\mathrm{NJP},c}^{\left(i\right)}\left(a_{j}\right),
\end{aligned}
\label{eq_decision}
\end{equation}
where $c=0,\cdots,MN-1$.
  \item \textbf{Stopping criteria}: The joint MP-MRC algorithm stops when at least one of the following conditions is satisfied.
  \begin{itemize}
    \item $\Upsilon^{\left(i\right)}=1$.
    \item $\Upsilon^{\left(i\right)}<\Upsilon^{\left(i^{*}\right)}-\epsilon$, where $i^{*}$ is the iteration index from $\left\{ 1,\cdots,\left(i-1\right)\right\}$ for which $\Upsilon^{\left(i^{*}\right)}$ is maximum.
    \item The maximum number of iterations is reached.
  \end{itemize}
\end{enumerate}

The computational complexity in one iteration of the MP-MRC algorithm presented here is $\mathcal{O}\left(PMNS|\mathbb{A}|\right)$, where $S=|\Phi_{pa}|=|\Psi_{pb}|$ represents the sparsity level of the channel matrix. Obviously, exploiting the sparsity of the channel is a key factor in reducing the computational complexity of the algorithm, and the corresponding simulation results are shown in the next section.

\section{Simulation Results}
\begin{table}
\centering
\caption{Simulation Parameters}
\begin{tabular}{|l|l|}
\hline
\textbf{Parameter}  & \textbf{Value}\tabularnewline
\hline
Carrier frequency & $4.0$ GHz\tabularnewline
\hline
Subcarrier spacing  & $15$ KHz\tabularnewline
\hline
Number of subcarriers ($M$) & $32$\tabularnewline
\hline
Number of OTFS symbols ($N$) & $16$\tabularnewline
\hline
Cyclic prefix duration   & $5$ $\mathrm{\mu s}$\tabularnewline
\hline
Modulation type & 16-QAM \tabularnewline
\hline
Number of BS antennas & 8, 32 \tabularnewline
\hline
Number of vehicle antennas & 1 \tabularnewline
\hline
\multirow{3}*{Channel Parameter}
&Tap number: $6$ \tabularnewline
\cline{2-2}
&Path in each tap: $8$ \tabularnewline
\cline{2-2}
&Maximum channel delay: $10$ \tabularnewline
\hline
Channel estimation & Ideal \tabularnewline
\hline
\end{tabular}
\label{tb}
\end{table}
This section shows the performance of the proposed OTFS based receiver with multi-antennas in the high-mobility vehicular communications. More specifically, we simulate the channel sparsity, the system error performance and the convergence performance of the MP-MRC iterative algorithm under different values of the receive antennas. The error performance of the OFDM based receiver is also presented as a comparison. The Jakes' channel model \cite{jakes1994microwave} between the moving vehicle and the BS is adopted. The antenna element spacing is $d=0.5\lambda$ and the angle information of the channel is assumed to be known at the receiver. Besides, in order to test the performance of the system under ultra-high mobility, for example, considering that the reflectors may include vehicles travelling in opposite directions, the maximum relative speed adopted here is 240 km/h \cite{guo2019high}. The other simulation parameters are shown in Table \ref{tb}.

\begin{figure}
  \centering
  \includegraphics[scale=0.52]{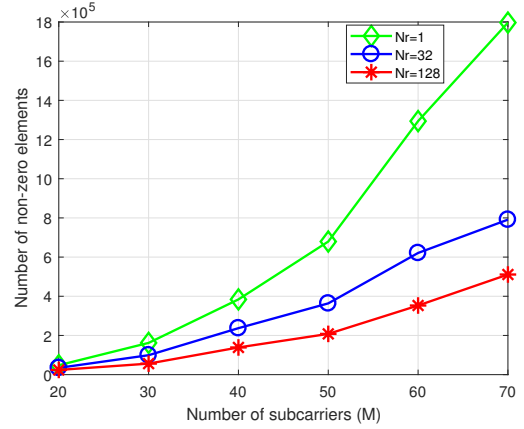}
  \caption{The number of non-zero elements of the channel matrix under different values of the receive antennas $N_{r}$.}
  \label{sparsity}
\end{figure}

Fig. \ref{sparsity} shows the channel sparsity of each beamforming branch in the delay-Doppler domain under different values of the receive antennas $N_{r}$. It is assumed that the elements in the channel matrix with modulo values less than $10^{-5}$ are zero elements. Consequently, the more zero elements in the channel matrix means the larger sparsity of the channel. For the easy of the illustration, we assume $N$ equals to $M$ in Fig. \ref{sparsity} and the normalized maximum DFO used here is $0.1$. It is observed from Fig. \ref{sparsity} that with the value of $N_{r}$ increases, the channel sparsity of each branch in the delay-Doppler domain also increases for the reason that more antennas can provide a higher spatial resolution to better separate the DFOs. Further, considering that the complexity in one iteration of the MP-MRC algorithm is $\mathcal{O}\left(PMNS|\mathbb{A}|\right)$, the enhanced channel sparsity after the beamforming network can considerably reduce the complexity of the algorithm.

\begin{figure}
  \centering
  \includegraphics[scale=0.435]{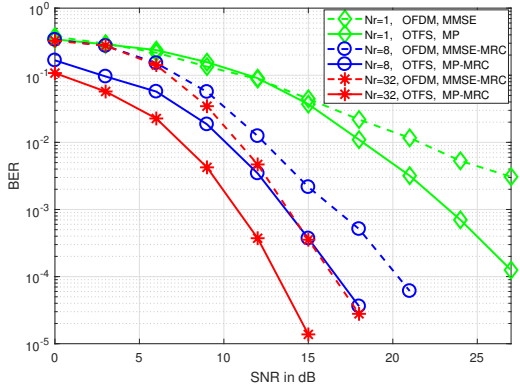}
  \caption{The BER performance comparison between the proposed OTFS based receiver and OFDM based receiver with different values of $N_{r}$.}
  \label{ber}
\end{figure}
\begin{figure}
  \centering
  \includegraphics[scale=0.52]{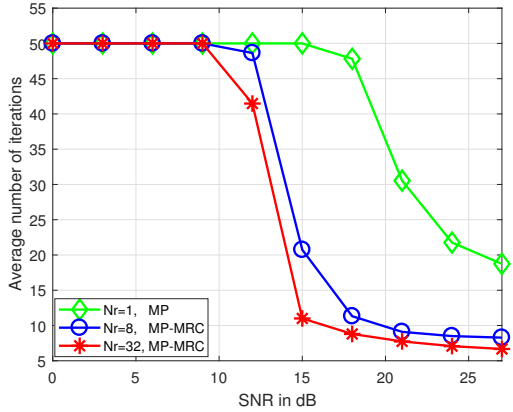}
  \caption{The convergence rate comparison between the proposed MP-MRC algorithm with multi-antennas and the MP algorithm with single antenna.}
  \label{convergence}
\end{figure}

In Fig. \ref{ber}, we compare the bit error rate (BER) performance between the proposed OTFS based receiver and the conventional OFDM based receiver with different values of $N_{r}$. We choose the damping factor $\Delta=0.5$ here and the minimum mean squared error-MRC (MMSE-MRC) algorithm is adopted for OFDM system in the case of $N_{r}>1$. From Fig. \ref{ber}, we can first observe that thanks to the multi-antenna diversity, the BER performance of both OTFS and OFDM systems improves as $N_{r}$ increases. More importantly, for any configurations of $N_{r}$, OTFS system achieves considerable gain compared to the OFDM system. For example, we can observe that the proposed OTFS based receiver outperforms the counterpart by approximately 4 dB at BER of $10^{-4}$ in the case of $N_{r}=8$ and $N_{r}=32$. The reason is that the ISFFT and SFFT operations of OTFS, together with the MP-MRC algorithm make all OTFS symbols in a frame experience a relatively constant channel gain, whereas in OFDM system, the overall performance is limited by the ICI caused by the DFOs.

In Fig. \ref{convergence}, the average number of iterations is simulated to analyze the convergence performance of the proposed joint MP-MRC iterative algorithm with multi-antennas and the MP algorithm with single antenna. From Fig. \ref{convergence}, we can observe that the convergence performance of our proposed MP-MRC algorithm is better than that of the MP algorithm for the reason that more independent channel information is utilized to improve the convergence performance in our proposed algorithm.

\section{Conclusion}
To combat the Doppler effect in the high-mobility V2X scenarios with large number of DFOs, we propose a novel OTFS based receiver scheme with multi-antennas in this paper. We show that the multiple DFOs associated with multipaths can be approximately separated with the high spatial resolution provided by the multi-antennas, which leads to an enhanced sparsity of the OTFS channel in the delay-Doppler domain and further reduces the complexity of the MP algorithm. Based on it, we propose a joint MP-MRC iterative detection for OTFS, where the integration of the MRC significantly improves the convergence performance of the algorithm and also obtains an excellent diversity gain. The simulation results show that the BER performance of the proposed OTFS based receiver is much better than that of the OFDM based receiver in the time-variant multipath channels, and the convergence performance of the proposed joint MP-MRC algorithm for OTFS also outperforms the conventional MP algorithm.

 \bibliographystyle{IEEEtran}
\bibliography{ciations,IEEEexample}

\begin{thebibliography}{10}
\providecommand{\url}[1]{#1}
\csname url@samestyle\endcsname
\providecommand{\newblock}{\relax}
\providecommand{\bibinfo}[2]{#2}
\providecommand{\BIBentrySTDinterwordspacing}{\spaceskip=0pt\relax}
\providecommand{\BIBentryALTinterwordstretchfactor}{4}
\providecommand{\BIBentryALTinterwordspacing}{\spaceskip=\fontdimen2\font plus
\BIBentryALTinterwordstretchfactor\fontdimen3\font minus
  \fontdimen4\font\relax}
\providecommand{\BIBforeignlanguage}[2]{{%
\expandafter\ifx\csname l@#1\endcsname\relax
\typeout{** WARNING: IEEEtran.bst: No hyphenation pattern has been}%
\typeout{** loaded for the language `#1'. Using the pattern for}%
\typeout{** the default language instead.}%
\else
\language=\csname l@#1\endcsname
\fi
#2}}
\providecommand{\BIBdecl}{\relax}
\BIBdecl

\bibitem{shah20185g}
S.~A.~A. Shah, E.~Ahmed, M.~Imran, and S.~Zeadally, ``{5G} for vehicular
  communications,'' \emph{IEEE Commun. Mag.}, vol.~56, no.~1, pp. 111--117,
  Jan. 2018.

\bibitem{cheng2015d2d}
X.~Cheng, L.~Yang, and X.~Shen, ``{D2D} for intelligent transportation systems:
  A feasibility study,'' \emph{IEEE Trans. Intell. Transp. Syst.}, vol.~16,
  no.~4, pp. 1784--1793, Aug. 2015.

\bibitem{jakes1994microwave}
W.~C. Jakes and D.~C. Cox, \emph{Microwave mobile communications}.\hskip 1em
  plus 0.5em minus 0.4em\relax Hoboken, NJ, USA: Wiley, 1994.

\bibitem{hadani2017otfs}
R.~{Hadani}, S.~{Rakib}, M.~{Tsatsanis}, A.~{Monk}, A.~J. {Goldsmith}, A.~F.
  {Molisch}, and R.~{Calderbank}, ``Orthogonal time frequency space
  modulation,'' in \emph{Proc. IEEE Wireless Commun. Netw. Conf.}, San
  Francisco, CA, Mar. 2017, pp. 1--6.

\bibitem{ramachandran2018mimo}
M.~K. Ramachandran and A.~Chockalingam, ``{MIMO-OTFS} in high-{Doppler} fading
  channels: Signal detection and channel estimation,'' in \emph{Proc. IEEE
  Global Commun. Conf.}, Abu Dhabi, United Arab Emirates, Dec. 2018, pp.
  206--212.

\bibitem{ding2019robust}
\BIBentryALTinterwordspacing
Z.~Ding, ``Robust beamforming design for {OTFS-NOMA},'' \emph{arXiv.org}, Oct.
  2019. [Online]. Available: \url{https://arxiv.org/abs/1910.14422}
\BIBentrySTDinterwordspacing

\bibitem{shen2019channel}
W.~Shen, L.~Dai, J.~An, P.~Fan, and R.~W. Heath, ``Channel estimation for
  orthogonal time frequency space {(OTFS)} massive {MIMO},'' \emph{IEEE Trans.
  Signal Process.}, vol.~67, no.~16, pp. 4204--4217, Aug. 2019.

\bibitem{ding2019otfs}
\BIBentryALTinterwordspacing
Z.~Ding, R.~Schober, P.~Fan, and H.~V. Poor, ``{OTFS-NOMA}: An efficient
  approach for exploiting heterogenous user mobility profiles,''
  \emph{arXiv.org}, Apr. 2019. [Online]. Available:
  \url{https://arxiv.org/abs/1904.02783}
\BIBentrySTDinterwordspacing

\bibitem{hadani2018arXiv}
\BIBentryALTinterwordspacing
R.~Hadani, S.~Rakib, S.~Kons, M.~Tsatsanis, A.~Monk, C.~Ibars, J.~Delfeld,
  Y.~Hebron, A.~J. Goldsmith, A.~F. Molisch \emph{et~al.}, ``Orthogonal time
  frequency space modulation,'' \emph{arXiv.org}, Aug. 2018. [Online].
  Available: \url{https://arxiv.org/abs/1808.00519}
\BIBentrySTDinterwordspacing

\bibitem{cheng2019low}
\BIBentryALTinterwordspacing
J.~Cheng, H.~Gao, W.~Xu, Z.~Bie, and Y.~Lu, ``Low-complexity linear equalizers
  for {OTFS} exploiting two-dimensional fast {Fourier} transform,''
  \emph{arXiv.org}, Sep. 2019. [Online]. Available:
  \url{https://arxiv.org/abs/1909.00524}
\BIBentrySTDinterwordspacing

\bibitem{xu2019low}
\BIBentryALTinterwordspacing
W.~Xu, T.~Zou, H.~Gao, Z.~Bie, Z.~Feng, and Z.~Ding, ``Low-complexity linear
  equalization for {OTFS} systems with rectangular waveforms,''
  \emph{arXiv.org}, Nov. 2019. [Online]. Available:
  \url{https://arxiv.org/abs/1911.08133}
\BIBentrySTDinterwordspacing

\bibitem{long2019low}
F.~Long, K.~Niu, C.~Dong, and J.~Lin, ``Low complexity iterative {LMMSE-PIC}
  equalizer for {OTFS},'' in \emph{Proc. IEEE Int. Conf. Commun.}, Shanghai,
  China, May 2019, pp. 1--6.

\bibitem{raviteja2018interference}
P.~Raviteja, K.~T. Phan, Y.~Hong, and E.~Viterbo, ``Interference cancellation
  and iterative detection for orthogonal time frequency space modulation,''
  \emph{IEEE Trans. Wireless Commun.}, vol.~17, no.~10, pp. 6501--6515, Oct.
  2018.

\bibitem{chizhik2004slowing}
D.~Chizhik, ``Slowing the time-fluctuating {MIMO} channel by beam forming,''
  \emph{IEEE Trans. Wireless Commun.}, vol.~3, no.~5, pp. 1554--1565, Oct.
  2004.

\bibitem{guo2017high}
W.~Guo, W.~Zhang, P.~Mu, and F.~Gao, ``High-mobility {OFDM} downlink
  transmission with large-scale antenna array,'' \emph{IEEE Trans. Veh.
  Technol.}, vol.~66, no.~9, pp. 8600--8604, Apr. 2017.

\bibitem{guo2019high}
W.~Guo, W.~Zhang, P.~Mu, F.~Gao, and H.~Lin, ``High-mobility wideband massive
  {MIMO} communications: {Doppler} compensation, analysis and scaling law,''
  \emph{IEEE Trans. Wireless Commun.}, vol.~18, no.~6, pp. 3177 -- 3191, Apr.
  2019.

\bibitem{rezazadehreyhani2018analysis}
A.~RezazadehReyhani, A.~Farhang, M.~Ji, R.~R. Chen, and B.~Farhang-Boroujeny,
  ``Analysis of discrete-time {MIMO OFDM-based} orthogonal time frequency space
  modulation,'' in \emph{Proc. IEEE Int. Conf. Commun.}, Kansas City, MO, Jul.
  2018, pp. 1--6.

\end{thebibliography}

\end{document}